\documentclass[conference]{IEEEtran}
\IEEEoverridecommandlockouts

\usepackage{cite}
\usepackage{amsmath,amssymb,amsfonts}
\usepackage{graphicx}
\usepackage{textcomp}
\usepackage{xcolor}
\def\BibTeX{{\rm B\kern-.05em{\sc i\kern-.025em b}\kern-.08em
    T\kern-.1667em\lower.7ex\hbox{E}\kern-.125emX}}

\usepackage{amsmath,amssymb,amsthm}
\usepackage{graphicx}
\usepackage{booktabs}
\usepackage{multirow}
\usepackage{array}
\usepackage{makecell}
\usepackage{algorithm}
\usepackage{algpseudocode}
\usepackage{xcolor}
\usepackage{url}
\usepackage[hidelinks]{hyperref}

\newtheorem{definition}{Definition}
\newtheorem{theorem}{Theorem}
\newtheorem{lemma}{Lemma}
\newtheorem{corollary}{Corollary}
\begin{document}

\pagestyle{plain}

\title{PriTabICL: Differentially Private Tabular
In-Context Learning with Task-Aligned Cluster-Wise Synthesis}

\author{
\IEEEauthorblockN{
Syed Ahzaz Shah\IEEEauthorrefmark{1},
Evan Bramblett\IEEEauthorrefmark{1},
Héber H. Arcolezi\IEEEauthorrefmark{2},
Chen Gong\IEEEauthorrefmark{3},
Qianqian Tong\IEEEauthorrefmark{1},
Xiaochen Li\IEEEauthorrefmark{1}
}
\IEEEauthorblockA{
\IEEEauthorrefmark{1}UNC Greensboro, Greensboro, USA
\quad
\IEEEauthorrefmark{2}ÉTS Montréal, Montréal, Canada
\\
\IEEEauthorrefmark{3}University of Virginia, Charlottesville, USA
}
\IEEEauthorblockA{
\texttt{\{shshah2,EDBRAMBLETT,q\_tong,x\_li12\}@uncg.edu}
\\
\texttt{heber.hwang-arcolezi@etsmtl.ca}
\quad
\texttt{fzv6en@virginia.edu}
}
}

\maketitle

\begin{abstract}
Tabular in-context learning (TabICL) enables large language models (LLMs) to predict target records using a handful of labeled demonstrations in their prompts. However, these demonstrations often contain sensitive personal information. Differential privacy (DP) offers principled protection, but noise-based methods introduce substantial noise, distorting the feature–label relationships essential for accurate prediction, especially as data dimensionality grows.

In this work, we study DP data synthesis for generating privacy-preserving TabICL demonstrations. We identify two limitations of directly applying general-purpose DP synthesis to entire datasets: modeling a global distribution can obscure local structure and minority patterns in heterogeneous data, while 
measuring broad attribute dependencies under
a limited privacy budget can reduce the accuracy
available for relationships most relevant to
prediction. More broadly, global statistical fidelity may fail to preserve heterogeneous structures needed by downstream predictive and generative models. To address these challenges, we propose PriTabICL, a general framework compatible with existing DP synthesis algorithms. It partitions records into DP clusters and synthesizes each independently, 
using a fixed feature--target workload to
prioritize prediction-relevant statistical
relationships. At inference, each query is routed to a DP cluster, and demonstrations are sampled from its synthetic dataset without additional privacy loss for the original records.
Experiments on four real-world datasets demonstrate the effectiveness of our methods across heterogeneous datasets. In particular, one of our instantiations, PriTabICL-PrivSyn, improves over global PrivSyn in 8 out of 12 settings and outperforms existing private TabICL approaches in nearly all evaluated settings. Open-source code is available\footnote{\url{https://github.com/ahzaz555/DPICL/tree/Pritabicl2026}}.

\end{abstract}


\section{Introduction}
\begin{figure}[t]
\centering
\includegraphics[width=\columnwidth]{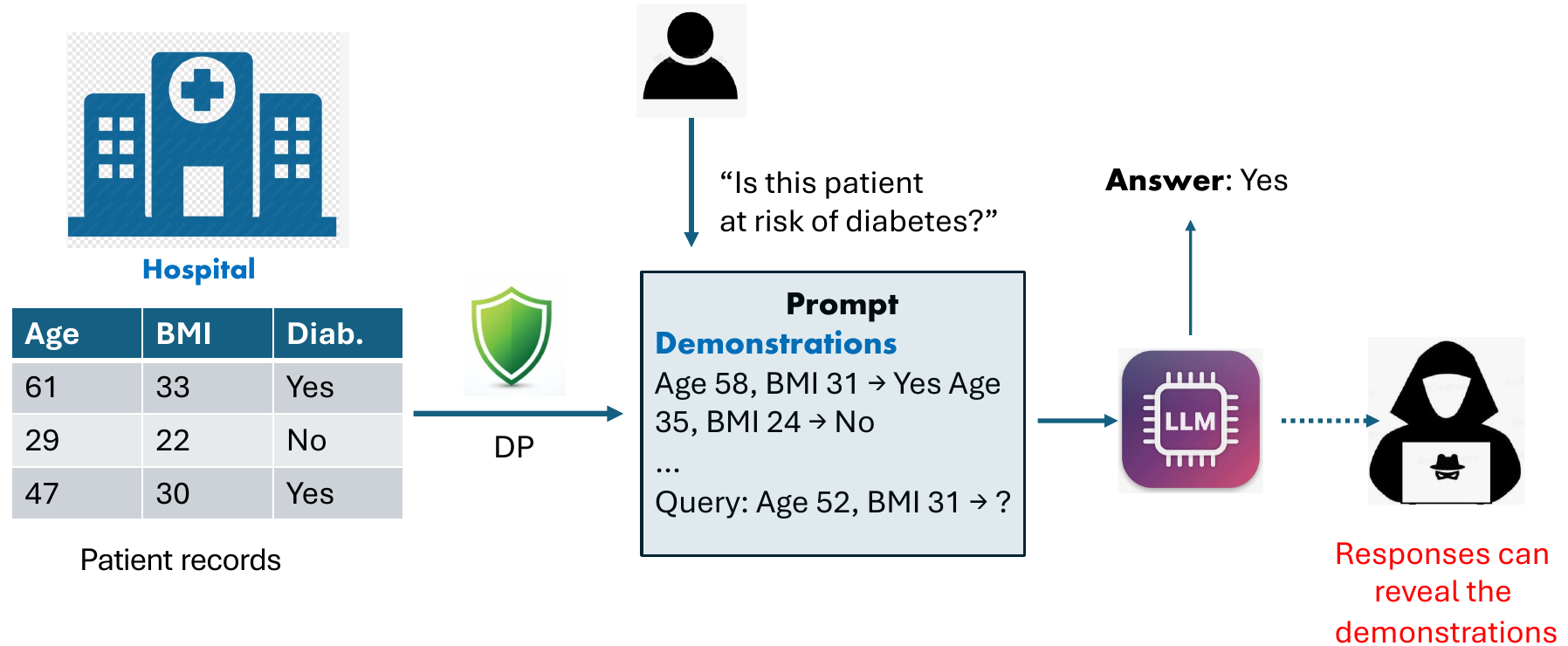}
\caption{A TabICL example and its privacy concern.}
\label{fig:scenario}
\vspace{-1em}
\end{figure}

In-context learning (ICL) enables large language models (LLMs) to perform tasks by reasoning over a few labeled examples provided in the prompt, without updating model parameters~\cite{brown2020language,min-etal-2022-rethinking}. Tabular in-context learning (TabICL) extends this paradigm to structured data: records  are serialized as text, a few labeled records (called \emph{demonstrations}) are included in the prompt, and the LLM predicts the label of an unlabeled record (the \emph{query})~\cite{hegselmann2023tabllm,han2024featllm,carey2024dptabicl}. This allows organizations to leverage their historical records through LLMs without training or deploying dedicated models. For example, a hospital could use patient records as demonstrations to support an LLM service that predicts whether a new patient is at risk of diabetes (Figure~\ref{fig:scenario}). However, exposing these records in prompts creates privacy risks: sensitive information may be extracted from model responses~\cite{zhang2024prompt} and an individual's presence in the prompt may be inferred through membership inference attacks~\cite{duan2023flocks}.

Differential privacy (DP)~\cite{dwork2014algorithmic} provides a rigorous mathematical framework for protecting individual records, ensuring that the output of a computation reveals little about whether any particular record is included in the dataset. DP can be instantiated under two common models: central DP (CDP), where a trusted curator adds noise to data statistics or computation results, and local DP (LDP)~\cite{kasiviswanathan2011can}, where each record is randomized before being shared. Common mechanisms include the Laplace and Gaussian mechanisms for numerical statistics and randomized response (RR) for categorical data~\cite{dwork2014algorithmic}. Existing work DP-TabICL~\cite{carey2024dptabicl} applies both CDP and LDP to protect TabICL demonstrations. However, independently perturbing individual attributes can destroy the correlations between features and labels that underpin accurate prediction. Moreover, as the number of attributes grows, the privacy budget must be distributed across more dimensions, potentially introducing increasingly substantial noise and degrading utility. Although existing high-dimensional DP publishing methods~\cite{ren2018textsf, chen2015differentially, xu2017dppro} exploit data structure to reduce perturbation error, the total amount of injected noise still grows linearly with dataset size.

An alternative is to generate synthetic data under DP. Rather than perturbing individual records or attributes, DP synthesis mechanisms learn privacy-protected statistical representations of the dataset and generate synthetic records from them~\cite{mckenna2021winning,mckenna2022aim,zhang2021privsyn,cai2021data,liu2021iterative, li2026heterofedsyn, harder2021dp}. Once a DP synthetic dataset is released, generating demonstrations from it incurs no additional privacy cost, regardless of the number of queries. Existing DP tabular synthesis methods broadly fall into two categories: model-based and graphical-model-based approaches. Recent benchmarks~\cite{chen2025benchmark} have systematically compared these methods and found that graphical-model-based approaches~\cite{mckenna2022aim,zhang2021privsyn, cai2021data} often achieve better utility than model-based approaches~\cite{liu2021iterative, harder2021dp}, despite their higher computational costs. These foundations make DP synthetic data a promising source of privacy-preserving TabICL demonstrations.

However, preserving global statistical properties by general-purpose DP synthesis may not be sufficient for TabICL, particularly when the private dataset contains heterogeneous subpopulations with distinct feature–label relationships. Modeling these subpopulations through a single global distribution aggregates their differences, potentially obscuring patterns specific to individual groups. Differential privacy can further exacerbate this information loss: privacy noise may smooth out fine-grained distributional differences, making subgroup-specific patterns, especially those of less prevalent groups, even harder to preserve. Thus, a synthetic dataset may retain the overall statistical properties while failing to capture local predictive relationships. This limitation is particularly important for TabICL, where predictions rely on only a few demonstrations whose informativeness depends on preserving the feature–label relationships relevant to each query.

A second challenge concerns how to allocate a limited privacy budget to preserve prediction-relevant information. General-purpose DP synthesizers typically aim to preserve a broad range of statistical dependencies among attributes to support diverse downstream tasks. However, preserving all these relationships incurs substantial privacy costs, while many of them may be irrelevant to TabICL, which primarily relies on the relationships between input features and target labels to construct informative demonstrations. Under a limited privacy budget, preserving these less relevant dependencies can compromise the preservation of prediction-relevant information required by TabICL.

\vspace{1mm}
\noindent\textbf{Our Contributions.}
We propose PriTabICL, a general framework for enabling privacy-preserving TabICL by integrating DP synthetic data into the inference pipeline. PriTabICL is compatible with existing DP data synthesis algorithms and performs all privacy-related data processing offline. The resulting DP synthetic data can then be used to support LLMs in answering queries without modifying the underlying models or introducing substantial inference overhead. To address the challenges of heterogeneous subpopulations and prediction-relevant information preservation, PriTabICL incorporates cluster-wise synthesis, feature–target statistical workloads, and query-based routing with balanced demonstration selection. Detailed contributions are as follows.

First, to address the loss of fine-grained distributional characteristics caused by global modeling, we introduce a cluster-wise DP synthesis strategy that captures local data distributions across heterogeneous subpopulations. Specifically, PriTabICL first partitions private records into groups based on data similarity, and then independently measures statistical properties and synthesizes each group to preserve its distinctive feature–target relationships. Although this strategy incurs additional privacy costs for clustering and group-wise synthesis, our analysis demonstrates that, on highly heterogeneous datasets, the information loss introduced by additional DP noise is outweighed by the loss incurred when modeling diverse subpopulations through a single global distribution.

Second, to address the inefficient allocation of limited privacy budgets to statistical relationships that are not essential for prediction, we introduce a task-aligned feature–target workload that prioritizes relationships between input features and the target label. Our workload provides a plug-and-play enhancement to existing DP synthesizers, requiring no redesign of their underlying mechanisms. We demonstrate its applicability through two instantiations built on two representative DP synthesizers, PrivSyn~\cite{zhang2021privsyn} and GEM~\cite{liu2021iterative}, enabling existing synthesis methods to better preserve prediction-relevant information under limited privacy budgets.

Finally, we conduct a comprehensive evaluation on four real-world datasets, comparing two PriTabICL instantiations, PriTabICL-PrivSyn and PriTabICL-GEM, against existing global DP synthesis baselines, a state-of-the-art privacy-preserving TabICL approach, and non-private method. Across the 12 evaluated settings, PriTabICL achieves the highest mean F1 among the compared private methods in 11 settings. In particular, PriTabICL-PrivSyn achieves an average F1 shortfall of only 0.013 relative to the best private method, with a maximum shortfall of 0.040. We further conduct component ablations and demonstration-level analyses to evaluate the contributions of task-aligned synthesis and investigate the importance of preserving label distinctions when constructing query-relevant demonstrations.

\section{Preliminaries}
\label{sec:preliminaries}

\subsection{Problem Statement}

Organizations can leverage their domain-specific data to enhance third-party LLM predictions through tabular in-context learning (TabICL), without training or fine-tuning the models. However, incorporating private records as labeled demonstrations into externally hosted LLM services may expose sensitive information to model providers or through model responses. In this paper, we consider this setting for binary classification. A trusted data provider holds a private labeled dataset \(D=\{(x_i,y_i)\}_{i=1}^{n}\), where \(x_i\in\mathcal{X}\) represents the features of record \(i\) and \(y_i\in\{0,1\}\) is its label. Given an unlabeled query \(x^*\), the objective is to construct privacy-preserving labeled demonstrations \(E(x^*)\) that allow the external LLM to predict its label \(\hat{y}\) while protecting individual records in \(D\).

\medskip
\noindent\textbf{Threat Model.}
Each record $(x_i,y_i)$, including both its features and its label, is one protected unit. The trusted curator may access the private dataset when running the privacy-preserving mechanisms. The LLM provider, downstream users, and anyone who sees the model's responses do not receive the original records; they may observe the constructed prompts, the outputs released by the curator, and the model responses, and may try to infer information about individuals in $D$. Because a response can reveal the demonstrations in its prompt~\cite{duan2023flocks,zhang2024prompt}, the demonstrations themselves must satisfy DP. Queries are external to $D$ and are not considered sensitive in our setting; protecting them is outside the scope of this work.

\subsection{Privacy Definitions}
\label{sec:privacy-definitions}

Differential privacy (DP) provides a mathematical guarantee
that limits how much the output of a randomized mechanism
can reveal about any individual record in a dataset.

\begin{definition}[$(\epsilon,\delta)$-Differential Privacy~\cite{dwork2014algorithmic}]
A randomized mechanism $\mathcal{M}$ satisfies
$(\epsilon,\delta)$-DP if, for any neighboring datasets
$D,D'$ and any measurable output set
$R\subseteq\operatorname{Range}(\mathcal{M})$,
\begin{equation*}
\Pr[\mathcal{M}(D)\in R]
\leq
e^\epsilon\Pr[\mathcal{M}(D')\in R]+\delta.
\end{equation*}
\end{definition}

We consider record-level add/remove adjacency, where
neighboring datasets differ by the addition or removal
of one record. Smaller values of $\epsilon$ and $\delta$
correspond to stronger privacy guarantees.

\medskip
\noindent\textbf{Zero-Concentrated Differential Privacy.}
Zero-concentrated differential privacy (zCDP)
provides an alternative privacy-accounting framework
for iterative mechanisms that perform multiple private
measurements, since their privacy costs can be
accumulated directly.

\begin{definition}[$\rho$-Zero-Concentrated Differential Privacy~\cite{bun2016concentrated}]
A randomized mechanism $\mathcal{M}$ satisfies
$\rho$-zCDP if, for every pair of neighboring datasets
$D,D'$ and every $\alpha>1$,
\begin{equation*}
D_\alpha\bigl(
\mathcal{M}(D)\|
\mathcal{M}(D')
\bigr)
\leq \rho\alpha,
\end{equation*}
where $D_\alpha(P\|Q)$ denotes the R\'enyi divergence
of order $\alpha$ between probability distributions
$P$ and $Q$.
\end{definition}

The Gaussian mechanism adds Gaussian noise with standard
deviation $\sigma$ to a function with $\ell_2$-sensitivity
$\Delta_2$, satisfying $\rho$-zCDP with
\begin{equation*}
\rho=\frac{\Delta_2^2}{2\sigma^2}.
\end{equation*}
This guarantee, together with the composition and conversion
properties of zCDP, follows from
Bun and Steinke~\cite{bun2016concentrated}.

For mechanisms satisfying $\rho_1,\ldots,\rho_r$-zCDP,
sequential composition gives
$(\sum_{j=1}^{r}\rho_j)$-zCDP.
A $\rho$-zCDP mechanism also satisfies
$(\epsilon,\delta)$-DP for any $0<\delta<1$, with
\begin{equation*}
\epsilon=
\rho+2\sqrt{\rho\ln(1/\delta)}.
\end{equation*}

\medskip
\noindent\textbf{Basic DP Mechanisms.}
The Laplace mechanism adds noise drawn from
$\operatorname{Lap}(\Delta_1/\epsilon)$,
where $\Delta_1$ is the $\ell_1$-sensitivity
of the released statistic, to achieve $\epsilon$-DP~\cite{dwork2014algorithmic}.
The Gaussian mechanism provides approximate DP
and supports the zCDP accounting described above.
These mechanisms are used in the private clustering
and synthesis procedures of PriTabICL.

\medskip
\noindent\textbf{Composition.}
Differential privacy satisfies sequential and parallel
composition ~\cite{dwork2014algorithmic}.
When multiple DP mechanisms access the same private
records, sequential composition accumulates their
privacy costs. In particular, composing mechanisms
with guarantees $(\epsilon_j,\delta_j)$ gives
$(\sum_j\epsilon_j,\sum_j\delta_j)$-DP.
When mechanisms operate on disjoint subsets and
each record influences only one mechanism, parallel
composition instead gives
$(\max_j\epsilon_j,\max_j\delta_j)$-DP.
Both properties are used in our end-to-end privacy analysis.

\medskip
\noindent\textbf{Post-Processing.}
Differential privacy is preserved under post-processing:
any computation performed using only DP outputs and
information independent of the private dataset
does not incur additional privacy
loss~\cite{dwork2014algorithmic}.

\subsection{DP Tabular Data Synthesis}
DP synthesis algorithms construct a synthetic dataset whose statistical properties approximate those of the original data \cite{zhang2021privsyn,liu2021iterative,chen2025benchmark}. Unlike row-wise perturbation, where each released record is obtained by randomizing an individual private record, data synthesis generates records from differentially private statistical information or a privately learned data distribution. Thus, the synthetic records are not perturbed copies of particular input records, while privacy is ensured through noise added during the private learning or measurement process.

Many tabular synthesis methods define the information to be preserved through statistical queries, which capture properties of the dataset such as attribute-value frequencies. A common example is a \emph{marginal}, which represents the frequency distribution over one or more attributes~\cite{zhang2021privsyn}. A collection of statistical queries targeted by a synthesis algorithm is called a \emph{workload}~\cite{liu2021iterative}. Since the privacy budget limits the accuracy of private measurements, selecting an appropriate workload is crucial for synthesis quality.

\medskip
\noindent\textbf{PrivSyn.}
PrivSyn~\cite{zhang2021privsyn} is a representative differentially private tabular data synthesis algorithm and has demonstrated strong performance in recent benchmark studies~\cite{chen2025benchmark}. PrivSyn first identifies informative low-dimensional marginals by estimating dependencies among attributes, allowing the limited privacy budget to focus on important statistical relationships. It then privately measures the selected marginals and applies consistency processing to reduce conflicts caused by noise. Finally, the Gradually Update Method (GUM) iteratively modifies an initial synthetic table to align it with the protected marginals while minimizing disruption to previously matched relationships across multiple attributes.

\medskip
\noindent\textbf{GEM.}
GEM~\cite{liu2021iterative} is a representative ML-based differentially private tabular data synthesis method that has demonstrated strong performance in recent benchmark studies~\cite{chen2025benchmark}. Instead of explicitly maintaining a synthetic table, GEM represents the synthetic distribution with a neural generator. It iteratively identifies workload queries with large errors, privately measures their answers on the private dataset, and updates the generator using the protected measurements. Synthetic records are then generated by sampling from the trained generator.

PrivSyn and GEM differ mainly in how they represent and optimize the synthetic distribution: PrivSyn gradually modifies a synthetic table to match selected private marginals, whereas GEM updates a neural generator to match an expanding set of privately measured query answers. We use both to show that PriTabICL does not depend on one synthesis approach.

\section{PriTabICL}
\label{sec:framework}
We now introduce PriTabICL, which aims to preserve local predictive patterns and prioritize statistical relationships relevant to classification. It first partitions private records into groups using DP clustering based on feature similarity. Each group is then synthesized separately using a feature–target workload, producing synthetic datasets that represent the measured relationships within their corresponding groups. Since this produces multiple synthetic datasets, PriTabICL routes each incoming query to a relevant DP cluster before selecting demonstrations for prediction.

\subsection{Overview}
\label{sec:overview}
\begin{figure}[t]
    \centering
    \includegraphics[width=\columnwidth]{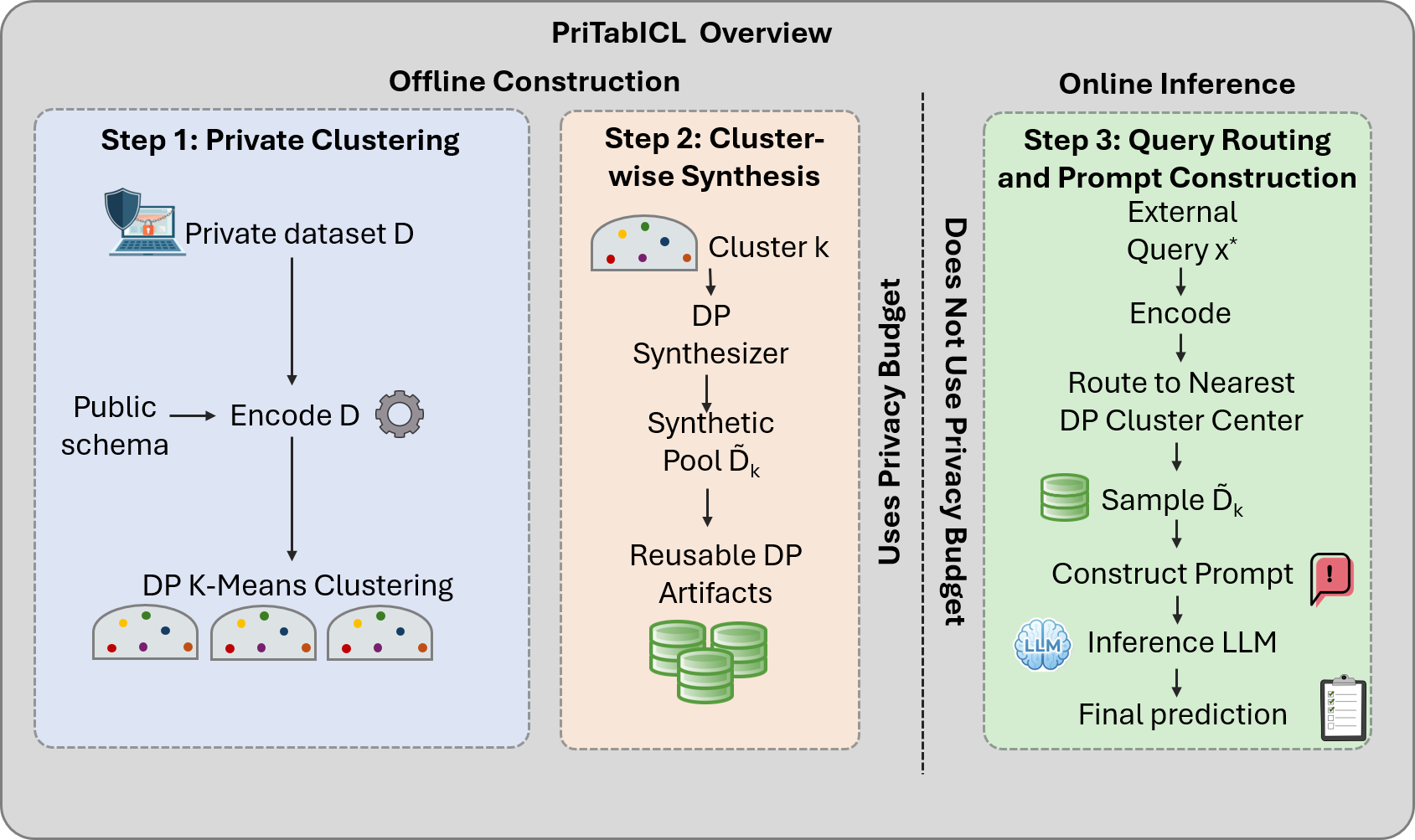}
        \caption{Overview of PriTabICL. Offline, DP clustering groups the private records and one DP synthetic dataset is built per cluster from statistics that relate each feature to the label. Online, each query is routed to its nearest DP center and receives label-balanced demonstrations drawn at random from that cluster's synthetic dataset.}
    \label{fig:framework}
\end{figure}

Figure~\ref{fig:framework} shows the workflow. An offline stage uses the private data once; an online stage answers queries using only its DP outputs. PriTabICL has three components, in the order they run.

\medskip
\noindent\textbf{DP Clustering.}
To capture predictive patterns that may be obscured
by global synthesis, PriTabICL first partitions
private records into groups of similar feature
profiles using DP clustering. This allows the
subsequent synthesis stage to model
feature--target relationships within each group,
rather than only across the entire dataset.
The resulting private record subsets are used
for cluster-wise synthesis, while the DP cluster
centers are retained for routing incoming queries.

\medskip
\noindent\textbf{Cluster-Wise DP Synthesis.}
PriTabICL synthesizes each private cluster separately to preserve feature--target relationships within individual groups rather than relying only on global statistics. To prioritize information useful for prediction, we use a feature--target workload that focuses the synthesis process on relationships
between input features and the target label.
This produces a DP synthetic dataset
$\widetilde{D}_k$ for each cluster, allowing
different groups to be represented by their
own synthetic records.

\medskip
\noindent\textbf{Query Routing and Cluster-Level Selection.}
Since PriTabICL produces a separate synthetic
dataset for each cluster, it must determine
which dataset to use for an incoming query.
PriTabICL routes the query to its nearest
DP cluster center using the same feature
representation as in clustering. It then
randomly samples labeled demonstrations
from the corresponding synthetic dataset,
balancing the labels when both are available.
The demonstrations and query are combined
into a prompt for LLM prediction. Because
this process uses only DP outputs and the
incoming query, it incurs no additional
privacy loss for the original records.

\begin{table}[t]
\centering
\caption{Notation.}
\label{tab:notation}
\scriptsize
\setlength{\tabcolsep}{3pt}
\begin{tabular}{@{}lp{0.64\columnwidth}@{}}
\toprule
Notation & Description \\
\midrule
$D$, $n$, $d$ & Private dataset, its size $n=|D|$ (private), number of non-target features \\
$N$ & Public size parameter, fixed in advance (not $|D|$) \\
$K$, $\widetilde{\mu}_k$ & Number of clusters, DP center of cluster $k$ \\
$D_k$, $\widetilde{D}_k$ & Private records assigned to cluster $k$, its synthetic dataset \\
$\mathcal{W}_Y$ & Feature--target workload $\{(X_j,Y)\}_{j=1}^{d}$ \\
$x^\ast$, $k^\ast$, $m$ & Query, the cluster it is routed to, number of demonstrations \\
$\epsilon$, $\epsilon_c$, $(\epsilon_s,\delta_s)$ & Total budget; budgets for clustering and for synthesis \\
\bottomrule
\end{tabular}
\end{table}

\subsection{Algorithm Design}
\label{sec:algorithm-design}

\begin{algorithm}[t]
\caption{PriTabICL}
\label{alg:pritabicl}
\begin{algorithmic}[1]
\Require \textbf{Offline:} private labeled dataset
$D$; public schema $S$; task description
$\mathcal{T}$; public size parameter $N$;
number of clusters $K$; privacy parameters
$\epsilon_c,\epsilon_s,\delta_s$
\Require \textbf{Online:} query $x^\ast$;
number of demonstrations $m$

\Statex \textbf{Offline construction}

\State Define the feature--target workload
$\mathcal{W}_Y=\{(X_j,Y)\}_{j=1}^{d}$

\State Encode the non-target features of
$D$ using the representation fixed by $S$

\State Run DP clustering with budget
$\epsilon_c$ to obtain
$\{\widetilde{\mu}_k\}_{k=1}^{K}$

\State Set $M\gets\lfloor N/K\rfloor$

\For{$k=1,\ldots,K$}
    \State Form the private subset $D_k$
    by assigning records to their nearest
    DP center
    \State Generate $M$ synthetic records
    $\widetilde{D}_k$ from $D_k$ using
    $\mathcal{W}_Y$ and
    $(\epsilon_s,\delta_s)$
\EndFor

\State Retain
$\{(\widetilde{\mu}_k,
\widetilde{D}_k)\}_{k=1}^{K}$

\Statex \textbf{Online inference}

\State Encode $x^\ast$ using the same
feature representation to obtain $z^\ast$

\State Route the query to
$k^\ast=\arg\min_k
\|z^\ast-\widetilde{\mu}_k\|_2^2$

\State Sample $m$ demonstrations
$\mathcal{E}(x^\ast)$ from
$\widetilde{D}_{k^\ast}$ using
label-balanced random selection

\State Construct the prompt using
$\mathcal{E}(x^\ast)$, $x^\ast$, $S$,
and $\mathcal{T}$

\State Obtain prediction $\hat{y}$
from the inference LLM

\State \Return $\hat{y}$
\end{algorithmic}
\end{algorithm}

Algorithm~\ref{alg:pritabicl} presents the
offline construction and online inference
procedures.

\medskip
\noindent\textbf{Offline Construction.}
Lines~1--4 define the feature--target
workload, encode the records, and perform
DP clustering. Lines~5--9 synthesize
the resulting private clusters and retain
each DP center with its synthetic dataset.This completes the privacy-preserving preprocessing stage, allowing subsequent queries to use the released synthetic data without accessing the original private records.

\medskip
\noindent\textbf{Online Inference.}
Lines~10--15 route each query to its
nearest DP center, construct demonstrations
from the corresponding synthetic dataset,
and obtain the LLM prediction.The selected cluster determines the source of demonstrations, which are sampled using the label-balanced random selection procedure.
Since inference uses only DP outputs and
public information, it incurs no additional
privacy loss for the original records.

\medskip
\noindent\textbf{An Example.}
Consider the Adult dataset with $K=3$ and a total
privacy budget $\epsilon=2$. PriTabICL uses
$\epsilon_c=0.6$ to privately group the records
into three clusters and $\epsilon_s=1.4$ to
synthesize each cluster separately using the
feature--target workload. This produces three
synthetic datasets that represent the measured
relationships within their respective groups.

For an incoming income-prediction query, PriTabICL
selects the nearest DP cluster center, say
cluster~2, and draws demonstrations from
$\widetilde{D}_2$. When both labels are available,
it randomly selects four synthetic records from
each income class to construct an eight-shot
prompt for the LLM. The query is answered using
the offline DP outputs without accessing the
original private records.

\section{Building Blocks}
\subsection{DP Clustering}
\label{sec:clustering}

A global feature--target distribution combines
the relationships of different subpopulations.
For a partition $D_1,\ldots,D_K$, the marginal
distribution of feature $X_j$ and target $Y$
can be expressed as

\begin{equation}
P_D(X_j,Y)
=
\sum_{k=1}^{K}
\frac{|D_k|}{|D|}
P_{D_k}(X_j,Y).
\label{eq:cluster-mixture}
\end{equation}

When feature--target relationships vary across
groups, their aggregation can obscure local
predictive patterns. Even an accurately
measured global marginal does not determine
the corresponding group-specific marginals.

PriTabICL therefore partitions records using
DP clustering of non-target features and
measures feature--target relationships within
each cluster. The target is excluded from
clustering because incoming queries are
unlabeled. The resulting DP centers are used
both to organize records for synthesis and
to route queries during inference.

\medskip
\noindent\textbf{Feature Representation.}
PriTabICL requires a bounded feature representation
to compare records with both numerical and
categorical attributes. Directly applying
Euclidean distance to raw attributes is
unsuitable because numerical features may
have different scales, while categorical
values have no meaningful numerical ordering.
We therefore construct a numerical
representation using the public data schema.

For each numerical feature $X_j$, the schema
specifies bounds $[L_j,U_j]$. We clip its value
to this interval and normalize it to $[0,1]$.
For each categorical feature, the schema
provides its domain $\Omega_j$, which we
encode using a one-hot vector scaled by
$1/\sqrt{|\Omega_j|}$. Formally, the encoded
coordinates for feature $X_j$ are

\begin{equation}
z_{i,B_j} =
\begin{cases}
\dfrac{
\operatorname{clip}(x_{ij},L_j,U_j)-L_j
}{U_j-L_j},
& X_j \text{ is numerical},\\[6pt]
\dfrac{\operatorname{onehot}(x_{ij})}
{\sqrt{|\Omega_j|}},
& X_j \text{ is categorical},
\end{cases}
\label{eq:feature-encoding}
\end{equation}
where $B_j$ denotes the coordinates
corresponding to feature $X_j$.

Under squared Euclidean distance, the
contribution of a numerical feature is at
most one, while a mismatch on a categorical
feature contributes $2/|\Omega_j|$.
The categorical scaling reduces the
influence of attributes with larger domains
using only public information. Since every
coordinate has a known bound, the
representation also provides the bounded
input required by DP clustering.

The target label is excluded from this
representation because the label of an
incoming query is unknown. Consequently,
the same encoding can be used to group
private records during offline construction
and to route unlabeled queries to the
resulting DP clusters.

\medskip
\noindent\textbf{DP Data Clustering.}
We use DPLloyd~\cite{su2016differentially},
implemented in diffprivlib~\cite{holohan2019diffprivlib},
to obtain private cluster centers.
Unlike ordinary $k$-means, DPLloyd updates
centers using privatized cluster counts
and coordinate sums. At iteration $t$,
coordinate $j$ of center $k$ is updated as

\begin{equation*}
\mu_{k,j}^{(t+1)}
=
\frac{\widetilde{s}_{k,j}^{(t)}}
{\widetilde{n}_k^{(t)}}.
\end{equation*}

Here, $\widetilde{n}_k^{(t)}$ and
$\widetilde{s}_{k,j}^{(t)}$ denote the
privatized count and coordinate sum.
The public feature bounds determine
the ranges used for noise calibration,
and $\epsilon_c$ is allocated across
the private updates of the clustering algorithm.

After clustering, each record is assigned
to its nearest DP center:

\begin{equation*}
D_k =
\left\{
(x_i,y_i)\in D:
k=\arg\min_{\ell}
\|z_i-\widetilde{\mu}_{\ell}\|_2^2
\right\}.
\end{equation*}

These assignments remain within the
trusted curator. For fixed DP centers,
the subsets are disjoint, allowing
cluster-wise synthesis to use parallel
privacy composition. The released DP
centers also support query routing
without accessing the original records.

\subsection{Cluster-Wise DP Synthesis}
\label{sec:synthesis}
PriTabICL synthesizes each private subset
$D_k$ independently to produce
$\widetilde{D}_k$, preserving
feature--target relationships at the
cluster level rather than only globally.
For fixed DP centers, the subsets are
disjoint, so each synthesizer can use the
full synthesis budget $(\epsilon_s,\delta_s)$
under parallel composition.

\medskip
\noindent\textbf{Feature--Target Workload.}
For classification, the joint distribution
$P_{D_k}(X_j,Y)$ captures how feature $X_j$
varies with the target within cluster $k$.
These relationships provide class-conditional
information that synthetic demonstrations
can convey to the LLM.

The connection can also be established
through the feature representation used
by PriTabICL. For an encoded query $z^\ast$
and an encoded record $Z$ from cluster $k$,
the expected squared distance to records
with label $y$ satisfies

\begin{equation}
\begin{aligned}
&\mathbb{E}\!\left[
\|z^\ast-Z\|_2^2
\mid Y=y,C=k
\right]\\
&\quad =
\sum_{j=1}^{d}
\mathbb{E}\!\left[
\|z^\ast_{B_j}-Z_{B_j}\|_2^2
\mid Y=y,C=k
\right].
\end{aligned}
\label{eq:class-conditional-distance}
\end{equation}

Each term depends only on the distribution
of feature $X_j$ within class $y$ of
cluster $k$. Consequently, the
feature--target marginals determine the
class-conditional expected distances and
their difference between the two classes.
Preserving these marginals therefore
retains a measurable aspect of the
class distinctions conveyed by labeled
demonstrations.

PriTabICL accordingly uses the
feature--target workload

\begin{equation}
\mathcal{W}_Y =
\{(X_j,Y):j=1,\ldots,d\}.
\label{eq:feature-target-workload}
\end{equation}

Rather than distributing synthesis effort
across all attribute pairs, this workload
prioritizes the relationships between each
input feature and the prediction target.

\medskip
\noindent\textbf{Effect on Measurement Noise.}
Consider a marginal-based synthesizer that
measures $r$ count histograms with a total
zCDP measurement budget $\rho_m$.
Assuming each histogram has
$\ell_2$-sensitivity one and the budget
is divided equally, the Gaussian noise
standard deviation is

\begin{equation*}
\sigma(r)=\sqrt{\frac{r}{2\rho_m}}.
\end{equation*}

For $d$ input features and one target,
a full pairwise workload contains
$d(d+1)/2$ marginals, whereas the
feature--target workload contains only $d$.
Under equal budget allocation, their
noise standard deviations satisfy

\begin{equation}
\frac{\sigma_{\mathrm{full}}}
{\sigma_{\mathrm{target}}}
=
\sqrt{\frac{d+1}{2}}.
\label{eq:workload-noise-ratio}
\end{equation}

Thus, focusing measurements on
feature--target relationships can reduce
the noise affecting prediction-relevant
statistics. This derivation assumes
uniform Gaussian measurement; practical
synthesizers may use different budget
allocations or adaptive selection.
PriTabICL incorporates the fixed workload
into PrivSyn and GEM as described next.

\medskip
\noindent\textbf{Instantiating PriTabICL with PrivSyn and GEM.}
We incorporate the feature--target workload
$\mathcal{W}_Y$ into two DP synthesizers,
PrivSyn and GEM. Both generate a synthetic
dataset $\widetilde{D}_k$ for each private
cluster $D_k$, but use different synthesis
mechanisms.

\medskip
\noindent\textit{PrivSyn.}
PrivSyn~\cite{zhang2021privsyn} constructs
synthetic records by measuring marginals
and updating a synthetic table to match them.
PriTabICL-PrivSyn replaces its data-dependent
marginal selection with the fixed workload
$\mathcal{W}_Y$. The original measurement
budget allocation, consistency processing,
and Gradually Update Method (GUM) are
retained; the budget allocated to private
workload selection is left unused.

Numerical features are discretized using
equal-width bins defined by the public
schema, while categorical features retain
their specified domains. After synthesis,
numerical bins are mapped back to values
using the public bin boundaries.

\medskip
\noindent\textit{GEM.}
GEM~\cite{liu2021iterative} uses a neural
generator updated from privately measured
statistical queries. PriTabICL-GEM replaces
its general pairwise workload with
$\mathcal{W}_Y$, while retaining GEM's
private query selection, measurement,
and generator optimization without altering these mechanisms.

Because GEM normalizes measurements by
dataset size and the cluster sizes are
private, we release a noisy size for
each cluster:

\begin{equation}
\widetilde{n}_k =
\max\left\{
1,\,
|D_k|+\operatorname{Lap}(1/\epsilon_n)
\right\}.
\label{eq:gem-noisy-size}
\end{equation}

This release consumes privacy budget
$\epsilon_n$; GEM uses the remaining
$(\epsilon_s-\epsilon_n,\delta_s)$.
Conditioned on $\widetilde{n}_k$,
the normalized-count sensitivity is
$1/\widetilde{n}_k$. The trained generator
produces $\widetilde{D}_k$.

\medskip
\noindent\textit{Other Synthesizers.}
PriTabICL can also use other DP synthesizers
that support independent synthesis of the
clusters. When configurable, their workloads
are set to $\mathcal{W}_Y$.

\medskip
\noindent\textbf{Measurement Precision within Clusters.}
For fixed DP centers, the disjoint clusters
allow each synthesizer to use the full
$(\epsilon_s,\delta_s)$ budget under parallel
composition. However, clustering reduces the
number of private records available to each
synthesizer within its assigned cluster.

For a count marginal measured with Gaussian
noise of standard deviation $\sigma$,
the relative noise scale in cluster $k$ is
$\sigma/n_k$, where $n_k=|D_k|$.
With approximately balanced clusters,
$n_k\approx n/K$, giving

\begin{equation*}
\frac{\sigma}{n_k}
\approx \frac{K\sigma}{n}.
\end{equation*}

Thus, cluster-wise synthesis trades finer
local modeling against increased relative
measurement noise. The feature--target
workload concentrates measurements on
prediction-relevant relationships but
does not eliminate this trade-off.

\medskip
\noindent\textbf{Synthetic Dataset Size.}
PriTabICL generates
$M=\lfloor N/K\rfloor$ synthetic records
per cluster, where $N$ is fixed publicly.
The released synthetic dataset sizes
therefore do not reveal private cluster
occupancy. The same synthesis procedure
is applied to empty clusters to avoid
revealing occupancy through missing
outputs or different output sizes.

\subsection{Query Routing and Balanced Demonstration Selection}
\label{sec:routing}

PriTabICL uses query similarity to select
a synthetic dataset, rather than individual
demonstrations. Because the feature--target
workload preserves class-conditional marginals
but not necessarily higher-order feature
dependencies, nearest-neighbor retrieval
from synthetic records may be unreliable.
We therefore route queries to a DP cluster
and sample demonstrations randomly within it.

\medskip
\noindent\textbf{Query Routing.}
Given a query $x^\ast$, we encode its features
as $z^\ast$ and select the nearest DP center:

\begin{equation*}
k^\ast =
\arg\min_{k=1,\ldots,K}
\|z^\ast-\widetilde{\mu}_k\|_2^2.
\end{equation*}

The selected center identifies the synthetic
dataset $\widetilde{D}_{k^\ast}$ without
accessing the original private records.

\medskip
\noindent\textbf{Balanced Demonstration Selection.}
For an even demonstration budget $m$,
PriTabICL randomly samples $m/2$ records
from each class when both labels are
available. This ensures that both classes
are represented rather than allowing
synthetic class proportions to determine
the prompt composition.

Let $\mathcal{E}_y$ denote demonstrations
sampled from class $y$. For any feature
value $v$, uniform sampling satisfies

\begin{equation*}
\begin{aligned}
&\mathbb{E}\left[
\widehat{P}_{\mathcal{E}_y}(X_j=v)
\mid \widetilde{D}_{k^\ast}
\right]\\
&\qquad =
P_{\widetilde{D}_{k^\ast}}
(X_j=v\mid Y=y).
\end{aligned}
\end{equation*}

Thus, the demonstrations preserve the
synthetic class-conditional feature
distributions in expectation without
requiring record-level nearest-neighbor
retrieval.

Sampling uses replacement when a class
has fewer records than required. If only
one class is available, all demonstrations
are drawn from that class. The selected
records are shuffled and combined with
the query and task description to form
the LLM prompt used for downstream prediction.

Since routing and demonstration selection
use only DP outputs and the incoming query,
they incur no additional privacy loss
for the original records.

\section{Theoretical Analysis}
\subsection{Privacy Analysis}
\label{sec:privacy-analysis}

We show that each component satisfies DP (Lemmas~\ref{lem:clustering} and~\ref{lem:synthesis}), prove the end-to-end guarantee (Theorem~\ref{thm:pritabicl-privacy}), and derive the guarantee of each version (Corollaries~\ref{cor:privsyn} and~\ref{cor:gem}), under record-level add/remove adjacency.

\begin{lemma}[DP clustering]
\label{lem:clustering}
DPLloyd with coordinate bounds taken from the schema, a number of iterations $T$ fixed from public values, and budget $\epsilon_c$ satisfies $\epsilon_c$-DP.
\end{lemma}
\begin{proof}
Each of the $T$ iterations releases a noisy count per cluster (sensitivity one) and noisy coordinate sums (bounded by the ranges in the schema) with budget $\epsilon_c/T$, so sequential composition gives $\epsilon_c$-DP~\cite{su2016differentially}; $T$ does not depend on $D$. The representation of ~\eqref{eq:feature-encoding} uses only the schema, so it introduces no additional access to the private data.
\end{proof}

\begin{lemma}[Cluster-wise synthesis]
\label{lem:synthesis}
Let $\mathcal{M}_s$ satisfy $(\epsilon_s,\delta_s)$-DP on every input. Conditional on the DP centers, applying $\mathcal{M}_s$ independently to $D_1,\ldots,D_K$ jointly satisfies $(\epsilon_s,\delta_s)$-DP.
\end{lemma}
\begin{proof}
For fixed centers, nearest-center assignment partitions $D$ into disjoint subsets, and the assignment of a record depends only on that record. Under add/remove adjacency, all shared records keep their assignment and the differing record affects exactly one subset. Each invocation is $(\epsilon_s,\delta_s)$-DP, including on an empty subset, so parallel composition gives the same guarantee for the joint release.
\end{proof}

\begin{theorem}[End-to-end privacy of PriTabICL]
\label{thm:pritabicl-privacy}
If DP clustering satisfies $\epsilon_c$-DP and cluster-wise synthesis satisfies $(\epsilon_s,\delta_s)$-DP, then PriTabICL satisfies
\begin{equation}
\left(\epsilon_c+\epsilon_s,\ \delta_s\right)\text{-DP}.
\label{eq:pritabicl-privacy}
\end{equation}
\end{theorem}
\begin{proof}
Synthesis takes the DP centers as input, so the two steps compose adaptively; by Lemmas~\ref{lem:clustering} and~\ref{lem:synthesis} and sequential composition, the release of the DP centers and all synthetic datasets is $(\epsilon_c+\epsilon_s,\delta_s)$-DP. Routing, demonstration selection, prompt construction, and LLM inference read only the query, public information, and these DP outputs, so by post-processing they incur no additional privacy loss, for any number of queries.
\end{proof}
\noindent 
The number of released synthetic records
is determined by the public parameters
$N$ and $K$, rather than the private
cluster sizes. The number of clustering
iterations is also fixed using public
parameters. When a synthesizer requires
a cluster size for its private measurements,
as in PriTabICL-GEM, the size is released
with differential privacy. Thus, the
published output sizes do not reveal
the number of private records assigned
to individual clusters to their relative sizes.

\begin{corollary}[PriTabICL-PrivSyn]
\label{cor:privsyn}
If each cluster-specific PrivSyn invocation is configured for $(\epsilon_s,\delta_s)$-DP, PriTabICL-PrivSyn satisfies ~\eqref{eq:pritabicl-privacy}.
\end{corollary}
\begin{proof}
The workload $\mathcal{W}_Y$ is fixed from the schema and the choice of target before any private measurement, so skipping PrivSyn's data-dependent marginal selection removes a private step and leaves its zCDP share unused. The one-way and feature--target marginals are measured with the Gaussian mechanism at PrivSyn's original shares of $\rho$, where $\rho$ is set so that $\rho$-zCDP implies $(\epsilon_s,\delta_s)$-DP; consistency processing and GUM are post-processing. The result follows from Theorem~\ref{thm:pritabicl-privacy} under this budget allocation.
\end{proof}

\begin{corollary}[PriTabICL-GEM]
\label{cor:gem}
If each cluster-specific GEM invocation is configured for $(\epsilon_s-\epsilon_n,\delta_s)$-DP given the released noisy size, PriTabICL-GEM satisfies ~\eqref{eq:pritabicl-privacy}.
\end{corollary}
\begin{proof}
The noisy sizes are a Laplace histogram over the disjoint clusters: one record changes one count by one, so their release is $\epsilon_n$-DP. Given $\widetilde{n}_k$, each answer of $\mathcal{W}_Y$ is a count divided by $\widetilde{n}_k$, which one record changes by at most $1/\widetilde{n}_k$, the sensitivity GEM uses; GEM's selection, measurement, and generator updates, and hence its zCDP accounting, are unchanged. By sequential composition, synthesis is $(\epsilon_s,\delta_s)$-DP, and Lemma~\ref{lem:synthesis} and Theorem~\ref{thm:pritabicl-privacy} give the result.
\end{proof}

\subsection{Complexity Analysis}
\label{sec:complexity}
\begin{theorem}[Computational Complexity of PriTabICL]
\label{the:computation}

Let $q$ be the encoded feature dimension,
$T$ the number of DPLloyd iterations,
and $T_{\mathrm{syn}}(n_k,d,M)$ the cost
of synthesizing $M$ records from a
cluster containing $n_k$ private records
and $d$ input features.
The offline computational cost is

\[
O\left(
TKnq+
\sum_{k=1}^{K}
T_{\mathrm{syn}}(n_k,d,M)
\right).
\]

Routing a query costs $O(Kq)$ in addition to demonstration sampling and LLM inference.
\end{theorem}
\begin{proof}
Encoding costs $O(nq)$, DPLloyd $O(TKnq)$, and synthesis $\sum_k T_{\mathrm{syn}}(n_k,d,M)$. Routing compares the query with $K$ centers in $O(Kq)$.
\end{proof}

\medskip
\noindent\textbf{Communication.} Each query sends only $m$ demonstrations, the query, and public prompt text to the LLM, as in standard $m$-shot TabICL; the private construction is offline and never delays a response.

\section{Experiments}
\label{sec:experiments}

We compare PriTabICL with existing private
TabICL approaches and global DP synthesis
baselines (Section~\ref{sec:exp-overall}),
isolate the contributions of the proposed
components (Section~\ref{sec:exp-components}),
and investigate when cluster-wise synthesis
improves classification
(Section~\ref{sec:exp-analysis}).
Additional findings on demonstration
selection and feature weighting are
discussed in Section~\ref{sec:discussion}.

\subsection{Experimental Setup}
\label{sec:experimental-setup}

\noindent\textbf{Datasets.}
We evaluate PriTabICL on four binary classification
datasets: Adult~\cite{adult_dataset}, Airline Passenger
Satisfaction~\cite{airline_survey}, Diabetes from the
2015 Behavioral Risk Factor Surveillance
System~\cite{brfss2015}, and Phishing
Websites~\cite{phishing_dataset}
(Table~\ref{tab:datasets}).
For Adult and Phishing, we use the splits provided
by SynMeter~\cite{du2024systematic}; Airline and
Diabetes are randomly divided into training,
validation, and test sets with a 60/20/20 split.
Training records with missing values are removed,
and queries are drawn exclusively from the test
sets.
Numerical bounds and categorical domains are
obtained from the public schema.

\begin{table}[h]
\centering
\setlength{\abovecaptionskip}{1ex} 
\caption{Datasets used in the evaluation.}
\label{tab:datasets}
\small
\setlength{\tabcolsep}{3.5pt}
\begin{tabular}{lrrrr}
\toprule
Dataset & \#Rows & \#Attr. & \#Num. & \#Cat. \\
\midrule
Adult & 19,313 & 14 & 6 & 8 \\
Airline & 77,675 & 22 & 4 & 18 \\
Diabetes & 152,208 & 21 & 21 & 0 \\
Phishing & 7,075 & 30 & 0 & 30 \\

\bottomrule
\end{tabular}
\end{table}

\begin{table*}[!t]
\centering
\caption{F1 (mean{\tiny$\pm$std} over 10 seeds, 100 class-balanced queries each). All methods except LDP-TabICL and GDP-TabICL draw four demonstrations of each class; Global PrivSyn and Global GEM draw them at random, ``+ kNN'' takes the four nearest synthetic records of each class. Bold: best private method. $\uparrow$/$\downarrow$: PriTabICL significantly better/worse than its own global baseline (paired $t$-test over seeds, Holm-corrected across the 12 settings, $p<0.05$).}
\label{tab:overall-f1}
\footnotesize
\setlength{\tabcolsep}{1pt}
\resizebox{\textwidth}{!}{%
\begin{tabular}{@{}l*{12}{c}@{}}
\toprule
& \multicolumn{3}{c}{Adult} & \multicolumn{3}{c}{Airline} & \multicolumn{3}{c}{Diabetes} & \multicolumn{3}{c}{Phishing} \\
\cmidrule(lr){2-4}\cmidrule(lr){5-7}\cmidrule(lr){8-10}\cmidrule(lr){11-13}
Method & $\epsilon{=}1$ & $\epsilon{=}2$ & $\epsilon{=}4$ & $\epsilon{=}1$ & $\epsilon{=}2$ & $\epsilon{=}4$ & $\epsilon{=}1$ & $\epsilon{=}2$ & $\epsilon{=}4$ & $\epsilon{=}1$ & $\epsilon{=}2$ & $\epsilon{=}4$ \\
\midrule
Non-private TabICL (reference) & 0.716{\tiny$\pm$0.030} & 0.716{\tiny$\pm$0.030} & 0.716{\tiny$\pm$0.030} & 0.724{\tiny$\pm$0.015} & 0.724{\tiny$\pm$0.015} & 0.724{\tiny$\pm$0.015} & 0.740{\tiny$\pm$0.019} & 0.740{\tiny$\pm$0.019} & 0.740{\tiny$\pm$0.019} & 0.727{\tiny$\pm$0.025} & 0.727{\tiny$\pm$0.025} & 0.727{\tiny$\pm$0.025} \\
Always predict positive & 0.667 & 0.667 & 0.667 & 0.667 & 0.667 & 0.667 & 0.667 & 0.667 & 0.667 & 0.667 & 0.667 & 0.667 \\
\midrule
LDP-TabICL & 0.658{\tiny$\pm$0.016} & 0.664{\tiny$\pm$0.015} & 0.670{\tiny$\pm$0.023} & 0.680{\tiny$\pm$0.025} & 0.677{\tiny$\pm$0.013} & 0.690{\tiny$\pm$0.024} & 0.725{\tiny$\pm$0.033} & 0.717{\tiny$\pm$0.037} & 0.725{\tiny$\pm$0.026} & 0.686{\tiny$\pm$0.010} & 0.677{\tiny$\pm$0.007} & 0.673{\tiny$\pm$0.007} \\
GDP-TabICL & \textbf{0.704}{\tiny$\pm$0.055} & 0.673{\tiny$\pm$0.079} & 0.654{\tiny$\pm$0.080} & 0.697{\tiny$\pm$0.020} & 0.704{\tiny$\pm$0.018} & 0.709{\tiny$\pm$0.009} & 0.672{\tiny$\pm$0.022} & 0.662{\tiny$\pm$0.016} & 0.669{\tiny$\pm$0.014} & 0.625{\tiny$\pm$0.027} & 0.612{\tiny$\pm$0.014} & 0.610{\tiny$\pm$0.013} \\
Global PrivSyn & 0.679{\tiny$\pm$0.036} & 0.676{\tiny$\pm$0.042} & 0.657{\tiny$\pm$0.037} & 0.732{\tiny$\pm$0.036} & 0.725{\tiny$\pm$0.024} & 0.722{\tiny$\pm$0.029} & 0.744{\tiny$\pm$0.013} & 0.734{\tiny$\pm$0.012} & 0.741{\tiny$\pm$0.015} & 0.695{\tiny$\pm$0.041} & 0.691{\tiny$\pm$0.014} & 0.721{\tiny$\pm$0.029} \\
Global PrivSyn + kNN & 0.513{\tiny$\pm$0.034} & 0.476{\tiny$\pm$0.055} & 0.491{\tiny$\pm$0.054} & 0.573{\tiny$\pm$0.052} & 0.565{\tiny$\pm$0.037} & 0.577{\tiny$\pm$0.041} & 0.728{\tiny$\pm$0.014} & 0.730{\tiny$\pm$0.014} & 0.735{\tiny$\pm$0.017} & 0.621{\tiny$\pm$0.070} & 0.598{\tiny$\pm$0.052} & 0.627{\tiny$\pm$0.045} \\
Global GEM & 0.638{\tiny$\pm$0.031} & 0.646{\tiny$\pm$0.031} & 0.647{\tiny$\pm$0.034} & 0.720{\tiny$\pm$0.022} & 0.721{\tiny$\pm$0.027} & 0.727{\tiny$\pm$0.040} & 0.755{\tiny$\pm$0.026} & 0.751{\tiny$\pm$0.031} & 0.750{\tiny$\pm$0.033} & 0.692{\tiny$\pm$0.054} & 0.720{\tiny$\pm$0.051} & 0.713{\tiny$\pm$0.037} \\
Global GEM + kNN & 0.521{\tiny$\pm$0.061} & 0.535{\tiny$\pm$0.059} & 0.540{\tiny$\pm$0.061} & 0.666{\tiny$\pm$0.059} & 0.671{\tiny$\pm$0.056} & 0.671{\tiny$\pm$0.052} & 0.727{\tiny$\pm$0.020} & 0.718{\tiny$\pm$0.015} & 0.726{\tiny$\pm$0.019} & 0.623{\tiny$\pm$0.047} & 0.624{\tiny$\pm$0.056} & 0.644{\tiny$\pm$0.046} \\
\midrule
PriTabICL-PrivSyn & 0.704{\tiny$\pm$0.064} & \textbf{0.693}{\tiny$\pm$0.040} & \textbf{0.689}{\tiny$\pm$0.034} & 0.730{\tiny$\pm$0.026} & 0.733{\tiny$\pm$0.035} & 0.735{\tiny$\pm$0.020} & 0.736{\tiny$\pm$0.022} & 0.730{\tiny$\pm$0.028} & 0.733{\tiny$\pm$0.015} & \textbf{0.727}{\tiny$\pm$0.029} & \textbf{0.751}{\tiny$\pm$0.026}\rlap{$^{\uparrow}$} & \textbf{0.744}{\tiny$\pm$0.052} \\
PriTabICL-GEM & 0.628{\tiny$\pm$0.044} & 0.626{\tiny$\pm$0.041} & 0.620{\tiny$\pm$0.054} & \textbf{0.743}{\tiny$\pm$0.037} & \textbf{0.748}{\tiny$\pm$0.038} & \textbf{0.748}{\tiny$\pm$0.039} & \textbf{0.774}{\tiny$\pm$0.022} & \textbf{0.770}{\tiny$\pm$0.022} & \textbf{0.767}{\tiny$\pm$0.025} & 0.687{\tiny$\pm$0.025} & 0.685{\tiny$\pm$0.020} & 0.683{\tiny$\pm$0.028} \\

\bottomrule
\end{tabular}%
}
\end{table*}

\medskip
\noindent\textbf{Baselines.}
We compare PriTabICL-PrivSyn and PriTabICL-GEM
against global PrivSyn~\cite{zhang2021privsyn}
and GEM~\cite{liu2021iterative}, which synthesize
the entire private dataset using their general
workloads and the full privacy budget.
Both global baselines use label-balanced random
demonstration selection. We also evaluate
balanced nearest-neighbor retrieval from the
same global synthetic datasets, selecting
the four nearest records of each class for each test query (``+ kNN'').

For existing private TabICL approaches, we
reproduce LDP-TabICL and GDP-TabICL
~\cite{carey2024dptabicl}. LDP-TabICL perturbs
individual records and selects the eight
nearest demonstrations, while GDP-TabICL
constructs demonstrations from private group
statistics. We retain their original selection
procedures rather than imposing label balancing;
GDP-TabICL is reproduced from its published
description because its code is unavailable.
Label-balanced random selection from the
original records serves as a non-private
reference.

\medskip
\noindent\textbf{Configuration.}
We evaluate privacy budgets
$\epsilon\in\{1,2,4\}$ with $\delta=10^{-5}$.
Unless stated otherwise, PriTabICL uses
$K=3$ clusters, $m=8$ demonstrations,
and a 30/70 privacy-budget split between
clustering and synthesis. PriTabICL-GEM
allocates $0.05\epsilon_s$ to releasing
noisy cluster sizes. The number of synthetic
records is determined by the public size
parameter $N$. Both synthesizers discretize
numerical features into eight equal-width
bins using public bounds.

\medskip

\noindent\textbf{Evaluation Protocol.}
We use Llama-2-13B-Chat for inference.
All methods are evaluated on the same 100 held-out
queries (50 per class), using ten matched random
seeds, a common prompt format, and the same output
parser. We report four classification metrics:

\begin{itemize}
    \item \textbf{F1-score (primary):}
    The harmonic mean of precision and recall,
    $F_1 = 2PR/(P+R)$. We report positive-class
    F1, capturing both false positive and
    false negative errors.

    \item \textbf{Precision:}
    The fraction of predicted positive cases
    that are correct, $P = TP/(TP+FP)$.

    \item \textbf{Recall:}
    The fraction of actual positive cases
    correctly identified, $R = TP/(TP+FN)$.

    \item \textbf{Accuracy:}
    The fraction of all correctly classified
    queries, $(TP+TN)/(TP+TN+FP+FN)$.
\end{itemize}

Here, TP, TN, FP, and FN denote true positives,
true negatives, false positives, and false
negatives, respectively. We report mean and
standard deviation over the ten matched seeds.
Accuracy, precision, and recall complement F1
by revealing different classification behaviors.

\noindent\textbf{Statistical Significance.}
For each dataset--budget setting, we compare
methods using paired $t$-tests on seed-level F1
scores. We apply Holm correction across the
12 settings for each method comparison and
consider adjusted $p<0.05$ significant.
Wilcoxon signed-rank tests on the 12 setting
means provide supplementary evidence, since
these settings are not fully independent.

\medskip
\noindent\textbf{Implementation.}
We run inference on NVIDIA L4 or A100 GPUs.
For PrivSyn, we correct the marginal-selection
noise calibration in the public implementation
to use its designated privacy-budget share,
following~\cite{chen2025benchmark}.
This correction affects the general-workload
baselines; PriTabICL-PrivSyn uses a fixed
workload without private marginal selection.
For GEM, we use its official implementation
with attribute domains specified by the schema.

\medskip
\noindent\textbf{Measures.}
Let $\mathrm{dist}(x,x')$ denote the squared
Euclidean distance between the feature
encodings of two records
(\eqref{eq:feature-encoding}).
For a query $x^\ast$ with true label
$y^\ast$, let
$\mathcal{E}^{+}$ and $\mathcal{E}^{-}$
denote demonstrations with matching
and different labels, respectively.

We measure query distance as the average
distance between $x^\ast$ and its
demonstrations, and diversity as the
average pairwise distance among
demonstrations. 
To examine whether demonstrations with
matching target labels are closer to the
query than those with different labels,
we define the label-separation gap as

\begin{equation}
\mathrm{gap}(x^\ast)
=
\overline{\mathrm{dist}}
(x^\ast,\mathcal{E}^{-})
-
\overline{\mathrm{dist}}
(x^\ast,\mathcal{E}^{+}),
\label{eq:class-gap}
\end{equation}

where $\overline{\mathrm{dist}}(x^\ast,S)$
is the average distance from the query
to records in $S$.
A positive gap indicates that demonstrations
with the same target label as the query
are closer, on average, than demonstrations
with a different target label.

We also measure the discrepancy between
the synthetic and original class-conditional
feature distributions:

\begin{equation}
\begin{aligned}
\operatorname{err}(\widetilde{D},D)
&=\frac{1}{2d}
\sum_y\sum_{j=1}^{d}\sum_v\\
&\quad\bigl|P_D(X_j=v\mid y)\\
&\qquad
-P_{\widetilde{D}}(X_j=v\mid y)\bigr|.
\end{aligned}
\label{eq:synthetic-error}
\end{equation}

This is the average $\ell_1$ discrepancy
across features and the two classes.
Numerical attributes are evaluated using
the same eight bins employed during
synthesis.

\subsection{Comparison of All Methods}
\label{sec:exp-overall}

To verify the effectiveness of PriTabICL, we compare all methods on four datasets under three privacy budgets. Table~\ref{tab:overall-f1} reports F1 for all methods.

\medskip
\noindent\textbf{Comparison with Global Synthesis.}
PriTabICL-PrivSyn improves on Global PrivSyn by 1.6 F1 points on average, is ahead in 8 of the 12 settings (Wilcoxon $p=0.027$), and is never significantly worse in any setting. The gain is concentrated on Adult (+2.5 points over its three budgets) and Phishing (+3.8), where it reaches 6.0 points at $\epsilon=2$, the one setting with a significant difference; on Airline and Diabetes the two are within one point of each other. PriTabICL-GEM behaves differently: it improves on Global GEM on Airline (+2.3 points over its three budgets) and Diabetes (+1.9) but falls behind on Adult ($-1.9$) and Phishing ($-2.3$), so that its average gain is zero (Wilcoxon $p=0.91$); no single setting differs significantly. 
These results show that the overall benefit
of PriTabICL varies across datasets and
synthesizers. Since PriTabICL changes both
the synthesis workload and the organization
of the private data, we examine their
individual contributions in
Section~\ref{sec:exp-components} and
further investigate the effect of
cluster-wise synthesis in
Section~\ref{sec:exp-analysis}.

\medskip
\noindent\textbf{Comparison with Existing Private TabICL.}
PriTabICL-PrivSyn exceeds LDP-TabICL in all 12 settings (Wilcoxon $p<0.001$; significantly in 6) and GDP-TabICL in 11 of 12 (Wilcoxon $p=0.001$; significantly in 8), by 3.9 and 6.0 F1 points on average; GDP-TabICL ties it only on Adult at $\epsilon=1$ (0.704). In accuracy, GDP-TabICL is the strongest private method on Adult and Airline: its group-summary demonstrations make the LLM predict the negative class more often, which raises accuracy on class-balanced queries but lowers recall and F1. PriTabICL-GEM exceeds LDP-TabICL and GDP-TabICL in 9 of 12 settings each. Across the private methods in Table~\ref{tab:overall-f1}, the best F1 comes from a version of PriTabICL in 11 of 12 settings, up to 10.9\% higher (relative) than the strongest existing private TabICL method (Phishing, $\epsilon=2$). One retrieval variant outside the table, which ignores label balance, is better on Phishing; we analyze it in Section~\ref{sec:discussion} when examining demonstration selection.

\begin{figure}[t]
\centering
\includegraphics[width=0.7\columnwidth]{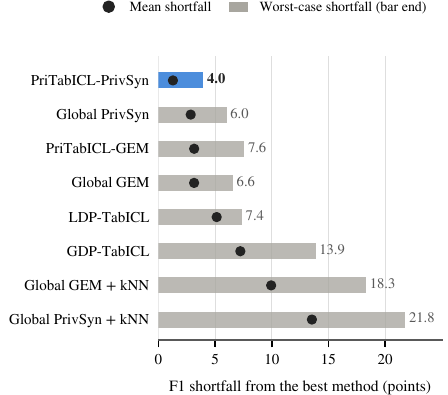}
\caption{PriTabICL-PrivSyn stays closest to the best private method. Each dot: F1 points below the best private method in one setting; bar: mean; right: worst case. Lower and narrower means a safer choice.}
\label{fig:safety}
\end{figure}

\medskip
\noindent\textbf{Consistency Across Datasets and Budgets.}
Each private baseline works well on some datasets and poorly on others: 
GDP-TabICL ties PriTabICL-PrivSyn for the
highest private F1 on Adult at $\epsilon=1$,
but falls 13.9 points behind the best
private method on Phishing and PriTabICL-GEM is the best on Airline and Diabetes but 7.6 behind on Adult (Figure~\ref{fig:safety}). 
Among the evaluated private methods,
PriTabICL-PrivSyn has the smallest
observed mean and maximum F1 shortfalls
relative to the best private method
in each setting: 1.3 and 4.0 points,
respectively, compared with 2.8 and
6.0 points for Global PrivSyn.
Its lowest observed F1 is 0.689,
the highest minimum among the evaluated
private methods.

\medskip
\noindent\textbf{Effect of the Privacy Budget.}
Increasing $\epsilon$ does not consistently
improve F1. For example, PriTabICL-PrivSyn
on Adult obtains 0.704, 0.693, and 0.689
at $\epsilon=1,2,4$, respectively.
Differences across privacy budgets are
generally small relative to the observed
run-to-run variation, suggesting that
downstream performance is influenced by
factors beyond the privacy noise alone,
including demonstration sampling and
LLM prediction behavior.

\medskip
\noindent\textbf{Comparison with the Non-Private Reference.}
PriTabICL-PrivSyn reaches or exceeds random selection from the original records on Airline and Phishing at every budget. On Diabetes, every method's F1 lies between 0.66 and 0.78, close to the 0.667 of always predicting the positive class, because the LLM predicts the positive class for most queries: PriTabICL-PrivSyn does so for 79\% of the class-balanced queries and reaches an accuracy of 0.66, whereas LDP-TabICL does so for 59\% and reaches 0.70 despite its lower F1. Differences on Diabetes should therefore be read with care.

\subsection{Contribution of Each Component}
\label{sec:exp-components}
\label{sec:exp-ablation}

\begin{table}[t]
\centering
\caption{Ablation and sensitivity at $\epsilon=2$
(F1, mean $\pm$ std over 10 seeds).
F--T denotes the feature--target workload.
Rows $K=2,4,5$ and budget 20/80, 40/60 vary
PriTabICL-PrivSyn defaults ($K=3$, 30/70).
$\uparrow$/$\downarrow$ indicate significantly
higher/lower F1 than the complete PriTabICL
configuration (paired $t$-test, Holm-corrected,
$p<0.05$).}
\label{tab:ablation}

\scriptsize
\setlength{\tabcolsep}{1.25pt}

\newcommand{\fscore}[3]{%
  \ensuremath{#1{\scriptstyle\pm #2}%
  \makebox[0.6em][c]{$\scriptstyle #3$}}%
}

\begin{tabular}{@{}lcccc@{}}
\toprule
Variant & Adult & Airline & Diabetes & Phishing \\
\midrule

\textbf{PriTabICL-PrivSyn}
& \fscore{0.693}{0.040}{}
& \fscore{0.733}{0.035}{}
& \fscore{0.730}{0.028}{}
& \fscore{0.751}{0.026}{} \\

\quad w/o clustering
& \fscore{0.728}{0.025}{}
& \fscore{0.729}{0.030}{}
& \fscore{0.731}{0.031}{}
& \fscore{0.698}{0.026}{\downarrow} \\

\quad w/o F--T workload
& \fscore{0.650}{0.025}{\downarrow}
& \fscore{0.711}{0.031}{}
& \fscore{0.725}{0.008}{}
& \fscore{0.726}{0.043}{} \\

\quad w/o both (global)
& \fscore{0.676}{0.042}{}
& \fscore{0.725}{0.024}{}
& \fscore{0.734}{0.012}{}
& \fscore{0.691}{0.014}{\downarrow} \\

\cmidrule(l){1-5}

\quad $K=2$
& \fscore{0.698}{0.039}{}
& \fscore{0.727}{0.022}{}
& \fscore{0.740}{0.019}{}
& \fscore{0.732}{0.029}{} \\

\quad $K=4$
& \fscore{0.693}{0.046}{}
& \fscore{0.716}{0.026}{}
& \fscore{0.736}{0.016}{}
& \fscore{0.745}{0.043}{} \\

\quad $K=5$
& \fscore{0.695}{0.036}{}
& \fscore{0.729}{0.022}{}
& \fscore{0.731}{0.017}{}
& \fscore{0.740}{0.037}{} \\

\quad budget 20/80
& \fscore{0.694}{0.060}{}
& \fscore{0.729}{0.031}{}
& \fscore{0.734}{0.021}{}
& \fscore{0.750}{0.040}{} \\

\quad budget 40/60
& \fscore{0.697}{0.057}{}
& \fscore{0.737}{0.019}{}
& \fscore{0.739}{0.018}{}
& \fscore{0.736}{0.033}{} \\

\midrule

\textbf{PriTabICL-GEM}
& \fscore{0.626}{0.041}{}
& \fscore{0.748}{0.038}{}
& \fscore{0.770}{0.022}{}
& \fscore{0.685}{0.020}{} \\

\quad w/o clustering
& \fscore{0.669}{0.042}{\uparrow}
& \fscore{0.719}{0.025}{\downarrow}
& \fscore{0.769}{0.017}{}
& \fscore{0.691}{0.028}{} \\

\quad w/o F--T workload
& \fscore{0.635}{0.067}{}
& \fscore{0.718}{0.029}{}
& \fscore{0.732}{0.020}{\downarrow}
& \fscore{0.720}{0.029}{\uparrow} \\

\quad w/o both (global)
& \fscore{0.646}{0.031}{}
& \fscore{0.721}{0.027}{}
& \fscore{0.751}{0.031}{}
& \fscore{0.720}{0.051}{} \\

\bottomrule
\end{tabular}
\end{table}

To investigate the contribution of the two
main components of PriTabICL, we conduct
ablation experiments at $\epsilon=2$
(Table~\ref{tab:ablation}).
We first replace the feature--target
workload with each synthesizer's general
workload while retaining cluster-wise
synthesis. We then remove clustering
while retaining the feature--target
workload, applying the synthesizer to
the complete private dataset.
All comparisons use the same
label-balanced demonstration selection.

\medskip
\noindent\textbf{Feature--Target Workload.}
Replacing the feature--target workload with
the general workload reduces PriTabICL-PrivSyn
F1 on all four datasets, by 2.4 points on
average and 4.3 points on Adult (significant).
Under global synthesis, the feature--target
workload provides its largest benefit on
Adult, increasing F1 from 0.676 to 0.728.

This benefit is consistent with concentrating
measurements on prediction-relevant
relationships: only 14 of Adult's 105
possible attribute pairs involve the target.
Such prioritization is particularly useful
within smaller clusters, where measurement
noise is larger relative to the sample size.

For GEM, the effect is mixed: +3.0 points
on Airline, +3.7 on Diabetes, and -3.5
on Phishing (the latter two significant),
averaging +0.6 points.

\medskip
\noindent\textbf{Clustering.}
With PrivSyn, clustering improves F1 by
5.3 points on Phishing (significant), has
little effect on Airline and Diabetes,
and reduces F1 by 3.4 points on Adult
($p=0.10$, not significant). The pattern
is similar under the general workload.
For GEM, clustering improves Airline
by 2.8 points but reduces Adult by
4.3 points, both significant.

\medskip
\noindent\textbf{Number of Clusters and Budget Split.}
Varying $K$ from its default of 3 to
2, 4, or 5, or changing the clustering/
synthesis budget split from 30/70 to
20/80 or 40/60, changes PriTabICL-PrivSyn
F1 by at most 1.9 points across the
evaluated datasets. None of these
differences is significant
(Table~\ref{tab:ablation}).

Overall, the feature--target workload
provides more consistent improvements
with PrivSyn, whereas clustering has
dataset-dependent effects. We investigate
these differences through synthetic-data
accuracy, query similarity, and label
separation in Section~\ref{sec:exp-analysis}.

\subsection{Analysis of Cluster-Wise Synthesis}
\label{sec:exp-analysis}

\begin{figure}[t]
\centering
\includegraphics[width=\columnwidth]
{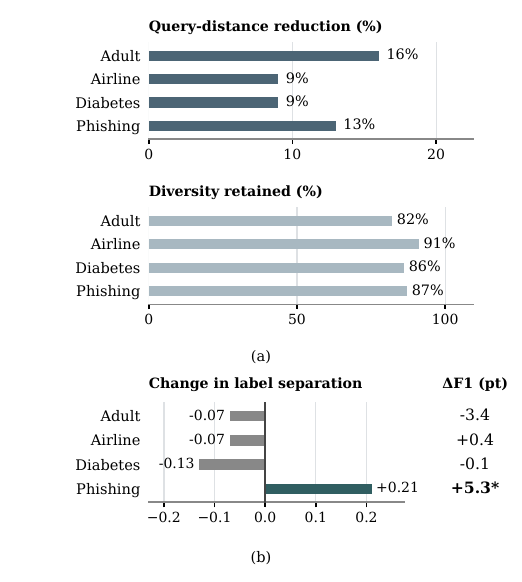}

\caption{Demonstration characteristics under
cluster-wise versus global PrivSyn at
$\epsilon=2$, using the same feature--target
workload. (a) Query-distance reduction and
diversity retained. (b) Changes in label
separation and F1 (points).
$^\ast$ denotes a significant F1 difference.}

\label{fig:demonstrations}
\end{figure}

The ablation results show that clustering
improves F1 on Phishing but provides little
benefit on the other datasets. To understand
this difference, we examine how cluster-wise
synthesis affects statistical accuracy and
the characteristics of demonstrations
selected for each query.

We compare PriTabICL-PrivSyn with global
PrivSyn using the same feature--target
workload at $\epsilon=2$, rebuilding the
synthetic datasets with matched seeds.
The analysis measures synthetic-data
accuracy, query similarity, diversity,
and class separation independently of
LLM inference. Statistics involving
private records are computed only for
experimental analysis and are not
released by PriTabICL.

\medskip
\noindent\textbf{Statistical Cost of Clustering.}
Cluster-wise synthetic datasets exhibit
greater feature-distribution error relative
to their corresponding private clusters
than global synthesis does relative to the
complete dataset (Table~\ref{tab:clusters}).
The difference reaches a factor of four
on Phishing.

This behavior is consistent with the
statistical cost of partitioning the data:
each cluster contains fewer records,
increasing the relative influence of
measurement noise. Consequently, the
improvements in downstream F1 cannot be
attributed to greater accuracy in the
measured feature distributions.

\begin{table}[t]
\centering
\caption{What DP clustering produces at $\epsilon=2$ (PrivSyn, 10 seeds). Clusters: non-empty clusters and the share of records in the largest. Rarer class: records of the cluster's rarer class behind the median query. Synthetic error: average class-conditional
feature-distribution discrepancy between
synthetic and real records, for cluster-wise
and global synthesis (lower is better).}
\label{tab:clusters}
\footnotesize
\setlength{\tabcolsep}{2.6pt}
\begin{tabular}{@{}lccccc@{}}
\toprule
Dataset & Clusters & \makecell{Largest\\share} & \makecell{Rarer\\class} & \makecell{Synthetic error\\cluster / global} \\
\midrule
Adult    & 1.8 & 0.81 & 4,415  & 0.087 / 0.073  \\
Airline  & 2.3 & 0.63 & 18,004 & 0.044 / 0.014 \\
Diabetes & 3.0 & 0.44 & 8,712  & 0.239 / 0.225  \\
Phishing & 3.0 & 0.73 & 2,308  & 0.013 / 0.003 \\
\bottomrule
\end{tabular}
\end{table}

\medskip
\noindent\textbf{Query Relevance and Class Separation.}
Cluster-wise synthesis brings demonstrations
9--16\% closer to their queries across
all four datasets while retaining
82--91\% of the diversity observed with
global synthesis
(Figure~\ref{fig:demonstrations}).
However, increased similarity alone does
not correspond to better classification
performance for every dataset.

On Phishing, the label separation gap increases
from 0.91 under global synthesis to
1.12 under cluster-wise synthesis,
accompanied by a significant improvement
of 5.3 F1 points. In contrast, the class
gap decreases on the other datasets.
For example, it decreases from 0.22
to 0.15 on Adult and from 0.14 to
0.01 on Diabetes, where clustering
provides no significant improvement in downstream classification F1.

These results suggest that the usefulness
of localized demonstrations depends not
only on their proximity to a query, but
also on whether the two classes remain
distinguishable within the selected
synthetic distribution. Clustering may
bring records closer together while
reducing the differences between their
class-conditional feature patterns.
The label separation gap captures this distinction
in the evaluated datasets and helps
explain why greater query similarity
does not necessarily improve F1. The gap measures average separation rather than the complete predictive structure of the synthetic records. It is therefore useful for interpreting the observed F1 changes, but does not guarantee better classification performance.

\medskip
\noindent\textbf{Cluster Structure.}
The effectiveness of localization also
depends on the clusters produced by the
DP clustering mechanism. On Adult at
$\epsilon=1$, DP clustering produces
only 1.6 nonempty clusters on average,
with 84\% of records assigned to the
largest cluster. The resulting synthesis
therefore offers limited separation
between subpopulations while still
incurring the clustering privacy cost.
In contrast, all three clusters remain
nonempty on Phishing.

Overall, the analysis identifies a
trade-off between statistical precision
and the predictive relevance of
demonstrations. Cluster-wise synthesis
can improve TabICL even when its measured
feature distributions are less accurate,
provided the localized demonstrations
retain useful distinctions between
classes. This explains the observed
benefit on Phishing and the more limited
gains on the other datasets.

\label{sec:exp-runtime}
\noindent\textbf{Runtime.}
On Phishing at $\epsilon=2$, mean offline
runtimes over three seeds are 39.77\,s
for Global PrivSyn versus 22.27\,s for
PriTabICL-PrivSyn, and 3.17\,s for
Global GEM versus 4.95\,s for
PriTabICL-GEM. PriTabICL timings include
DP clustering and cluster-wise synthesis;
global timings include synthesis only.
LLM inference is excluded.

\section{Discussion}
\label{sec:discussion}
\noindent\textbf{Demonstration Selection.}
Although nearest-neighbor retrieval is
commonly used to select demonstrations,
our experiments show that it does not
necessarily benefit TabICL with DP
synthetic records. Compared with
label-balanced random selection from
the same synthetic dataset, balanced
retrieval reduces F1 by 10.7 points
with PrivSyn and 6.8 points with GEM
on average across the 12 settings.

Real-record controls help explain this
behavior. On Airline and Phishing,
retrieving the nearest records regardless
of label benefits from the local majority
label, but forcing equal numbers of
demonstrations from each class removes
much of this advantage. Balanced
retrieval also underperforms random
sampling on the original records. This control indicates that the problem is not caused by DP synthesis alone. Even with real records, forcing a balanced set of nearest neighbors can remove useful information about the local class distribution.

The disadvantage becomes more pronounced
on several DP synthetic datasets.
This is consistent with marginal-based
synthesis preserving class-conditional
feature distributions more directly
than the higher-order relationships
on which record-level proximity relies.
Nevertheless, unbalanced retrieval from
DP synthetic records still outperforms
PriTabICL on Phishing at $\epsilon=2$
(0.789 versus 0.751 F1). Thus,
the most suitable demonstration-selection
strategy depends on the predictive
information retained by synthesis.

\medskip
\noindent\textbf{Feature Weighting.}
Our cluster-wise analysis also motivates
weighting features according to their
predictive relevance. We evaluated two
approaches using Qwen2.5-14B-Instruct
~\cite{qwen2024qwen25} and DP
class-conditional statistics computed
from a disjoint 10\% subset of records.
One uses noisy numerical means and
categorical modes, while the other uses
full noisy feature histograms for each target class.

Summary-based weighting reduces F1 by
0.6 points on average relative to equal
relevance weights. Full histograms
reduce this loss but do not provide a
statistically significant improvement
over equal weights or PriTabICL using
all records without learned weighting.
We therefore retain the simpler
schema-scaled feature representation.

\section{Related Work}
\label{sec:related_work}
\noindent\textbf{Tabular In-Context Learning.}
In-context learning (ICL) enables large language models to perform tasks from examples provided directly in the prompt without updating model parameters
~\cite{brown2020language,min-etal-2022-rethinking}. This capability has been extended to tabular prediction by representing structured records as text and using
labeled records as demonstrations
~\cite{hegselmann2023tabllm,han2024featllm}. Related work has further explored feature-oriented representations and LLM-based tabular generation and
synthesis
~\cite{wang2019tafenet,fang2025tabgen,fang2024large,
DBLP:conf/nips/WangFDCHAXW24}. 

The use of sensitive records as demonstrations has motivated
privacy-preserving ICL. Existing approaches construct private demonstrations or prompts through differentially private generation and aggregation
~\cite{tang2024privacy,gao2025data}. In the tabular setting,
DP-TabICL~\cite{carey2024dptabicl} considers both local perturbation of
individual records and central-DP construction from aggregate statistics. More recently, TabPATE~\cite{wahdany2026tabpate} splits the private data among teachers, labels generated records by their noisy vote, and uses the labeled records as one context shared by all predictions of a tabular foundation model (TabPFN). PriTabICL instead builds DP synthetic data for the prediction task and selects demonstrations for each query by cluster. Because TabPATE targets a tabular foundation model rather than an LLM and has no public implementation, our experiments compare with DP-TabICL, which shares our LLM setting.

\medskip
\noindent\textbf{Differentially Private Tabular Synthesis.}
Differentially private tabular synthesis has been studied through a broad range of statistical and graphical approaches based on marginal
measurements, dependency structures and probabilistic models
~\cite{DBLP:conf/sigmod/ZhangCPSX14,DBLP:conf/icml/McKennaSM19,
mckenna2021winning,mckenna2022aim,zhang2021privsyn,cai2021data}.
Methods such as PrivBayes, Private-PGM, MST, AIM, PrivSyn, and PrivMRF differ in how they identify, measure, and fit useful statistical relationships, but share the goal of producing a reusable synthetic approximation of the private
data. In particular, recent methods such as AIM and PrivSyn place emphasis on selecting informative measurements under a limited privacy
budget.

A complementary line of work uses neural generative models for private data
generation
~\cite{xie2018differentially,jordon2018pate,liu2021iterative}. Rather than explicitly fitting a graphical representation or synthetic table, these
methods learn a generator from privacy-protected information. GEM
~\cite{liu2021iterative}, for example, iteratively measures selected workload queries and optimizes a neural generator against their protected answers.
Recent benchmarking shows that statistical and neural synthesis mechanisms can exhibit different utility characteristics across datasets and workloads
~\cite{chen2025benchmark}. We therefore use PrivSyn and GEM as representative but substantially different synthesizers.
Most prior work targets general-purpose data release, aiming to preserve statistical structure for diverse downstream analyses. In TabICL, however, synthetic records serve as labeled demonstrations, and each prediction uses only a small, query-specific subset. Consequently, preserving global statistics alone may not retain the local structure and feature--target relationships needed for effective prediction. Although prior work has explored partitioning data before synthesis to model more homogeneous subsets~\cite{benkhelif2017coclustering,bowen2021steps}, such partitions are not necessarily aligned with downstream prediction tasks.

\section{Conclusion}
We proposed PriTabICL, a DP synthesis
framework that combines cluster-wise
modeling with a task-aligned
feature--target workload to construct
privacy-preserving TabICL demonstrations.
Experiments with PrivSyn and GEM show
that the targeted workload provides
consistent benefits for PrivSyn, while
the effectiveness of clustering depends
on the preservation of useful local
label distinctions.

Future work will investigate adapting
the number of clusters and privacy-budget
allocation to the characteristics of
the dataset using DP measurements,
avoiding unnecessary partitioning when
clustering provides limited benefits.
We also plan to extend the workload
beyond pairwise feature--target statistics
to capture higher-order label interactions
and explore demonstration-selection
strategies that better exploit the
predictive structure retained by
synthetic data.

\section*{Acknowledgment of AI-Generated Content}
The authors used AI-assisted tools, including ChatGPT (OpenAI) and Claude (Anthropic), to support manuscript review, improve clarity and organization, and assist with review of coding and data analysis. All research decisions, experimental results, and conclusions were reviewed and validated by the authors. Llama-2-13B-Chat and Qwen2.5-14B-Instruct are part of the evaluated pipelines.

\bibliographystyle{IEEEtran}
\bibliography{sample.bib}

\end{document}